# Vendor-Agnostic Pseudo-continuous Arterial Spin Labeling Using Pulseq: Cross-Vendor Quantitative Evaluation

**Jianing Zhang[1,2], Jun Wang[3], Joseph G. Woods[4], Shaoyou Ye[2], Shuai Ming[5], Hongwei Li[6], Wei Hu[1], Wei Wei[5], Xiaohu Li[3], Thomas W. Okell[7,8], Yang Ji[1,2]**

[1] Department of Neurology, The First Affiliated Hospital of USTC, University of Science and Technology of China, Hefei, China

[2] Department of Electronic Engineering and Information Science, School of Information Science and Technology, University of Science and Technology of China, Hefei, China

[3] Department of Radiology, The First Affiliated Hospital of Anhui Medical University, Hefei, China

[4] Department of Diagnostic, Interventional and Pediatric Radiology, Inselspital, Bern University Hospital, University of Bern, Bern, Switzerland

[5] Department of Radiology, The First Affiliated Hospital of USTC, University of Science and Technology of China, Hefei, China

[6] Department of Radiology, School of Medicine, University of Washington, Seattle, WA, USA

[7] The Podium Institute for Sports Medicine and Technology, Department of Engineering Science, University of Oxford, Oxford, United Kingdom

[8] Oxford Centre for Integrative Neuroimaging, FMRIB, Nuffield Department of Clinical Neurosciences, University of Oxford, Oxford, United Kingdom

Correspondence
Yang Ji, Ph.D., Department of Electronic Engineering and Information Science, School of Information Science and Technology, University of Science and Technology of China, Hefei, China.
Email: jiyang@ustc.edu.cn

Abstract word count: 280
Manuscript word count: ~4740
Figure count: 10
Table count: 0

## Acknowledgments

This work was supported by the National Natural Science Foundation of China (NSFC; grant 62401535 to Y.J.), the Wellcome Trust and the Royal Society (Sir Henry Dale Fellowship 220204/Z/20/Z to T.W.O.) and the Podium Institute for Sports Medicine and Technology at the University of Oxford (T.W.O.). The Oxford Centre for Integrative Neuroimaging (OxCIN) was supported by core funding from the Wellcome Trust (203139/Z/16/Z) with additional support from the NIHR Oxford Biomedical Research Centre (NIHR203311) and the Oxford Health Biomedical Research Centre (NIHR203316). For the purpose of open access, the author has applied a CC BY public

## Abstract

**Purpose:** Inter-scanner variability in cerebral blood flow (CBF) quantification hampers the reproducibility of multi-center ASL studies. This study aimed to develop and validate a vendor-agnostic pseudo-continuous ASL (PCASL) sequence using the open-source Pulseq framework to improve inter-scanner reproducibility of CBF measurements, and to demonstrate extensibility to advanced perfusion imaging.

**Methods:** A vendor-independent PCASL sequence was implemented using Pulseq and deployed on 3T scanners from three vendors (Siemens, GE, and United Imaging). Seven healthy volunteers underwent single-post-labeling delay (PLD) PCASL acquisitions across all platforms, including test–retest scans for intra-scanner evaluation. The in-house adapted vendor-based PCASL implementation on Siemens (Siemens-Vendor) served as reference for comparisons with Pulseq implementations. Reproducibility was assessed using within-subject coefficient of variation ($CV_{within}$), root mean square error (RMSE), Pearson correlation of regional CBF, and voxel-wise variation ratio mapping. Multi-PLD PCASL and planning-free VE-PCASL were implemented as proof-of-concept extensions.

**Results:** Mean gray matter CBF showed close agreement across all implementations (range: 52.25–53.64 mL/100g/min). Test–retest GM $CV_{within}$ remained below 4% for all platforms, with Siemens-Pulseq achieving the lowest value (2.65%), outperforming the native sequence (3.64%). GM RMSE between Siemens-Vendor and Pulseq implementation (1.64–1.96 mL/100g/min) was comparable to test–retest RMSE (1.97–2.68 mL/100g/min), with all inter-scanner $CV_{within}$ below 3%. Voxel-wise variation ratio maps showed distributions centered near 100%, suggesting that cross-vendor variability was comparable in magnitude to physiological baseline fluctuations. Multi-PLD and VE-PCASL modules executed successfully across all three platforms, yielding consistent CBF/ATT maps and vascular territory delineations.

**Conclusion:** The Pulseq-based PCASL sequence enables vendor-agnostic perfusion imaging with low inter-scanner variability. This open-source approach provides a reproducible and standardized solution for multi-center ASL studies and lowers the barrier for deploying advanced ASL techniques in clinical research.

## 1 Introduction

Arterial spin labeling (ASL) is a non-invasive MRI technique that magnetically labels arterial blood water to generate perfusion-weighted images of the brain(1,2). Among the available ASL approaches, pseudo-continuous ASL (PCASL)(3,4) is widely used in clinical and research applications because of its favorable signal-to-noise ratio (SNR) and compatibility with standard clinical MRI systems (5,6). Quantitative measures derived from PCASL, including cerebral blood flow (CBF) and arterial transit time (ATT), are increasingly used in studies of neurological disorders, including stroke, cerebrovascular disease, and traumatic brain injury (7-12). In these studies, multi-center data collection can provide larger sample sizes and improve the generalizability, statistical power, and reliability of quantitative analyses(13-15).

However, inter-scanner variability remains an important limitation for quantitative ASL in multi-center studies(5,16,17). Although global gray matter CBF is broadly reproducible, regional perfusion and white matter CBF exhibit considerable variability across scanners and sites(16,18). Part of this variability originates from differences in sequence implementation. Vendor-specific PCASL sequences are typically closed-source and may differ in labeling strategies, event timing, readout schemes, and background suppression approaches(19-21). These differences, together with hardware-dependent effects, can introduce systematic variation in measured perfusion and complicate the pooling and interpretation of multi-center ASL data (22). In addition, implementing and modifying advanced ASL methods separately within proprietary programming environments is time-consuming and requires vendor-specific expertise, which can hinder the translation of methodological developments across scanner platforms. Therefore, methods that promote data consistency across sites are necessary to enable robust pooled analyses(23).

An important distinction in multi-center ASL is therefore between harmonizing nominal acquisition parameters and harmonizing the underlying acquisition itself. Parameters such as labeling duration, repetition time (TR), and echo time (TE) can be matched through the scanner user interface, while the sequence implementations underlying these parameters may remain different across vendors (24,25). Indeed, nominally matched PCASL protocols can produce different quantitative CBF measurements across 3T MRI systems(16,17). Differences in readout schemes and background suppression strategies can affect SNR, tissue contrast, and signal stability (26), while differences in RF pulse design for blood labeling can alter labeling efficiency and consequently quantitative perfusion estimates(27). Developing a vendor-agnostic PCASL sequence is therefore essential to ensure that internal sequence strategies and pulse designs remain identical across platforms, reducing inter-scanner variability and promoting ASL standardization(28). A sequence that is defined independently of a specific vendor and executed using the same sequence logic across platforms therefore provides a potential approach to reducing acquisition-level sources of inter-scanner variability

Several vendor-neutral frameworks for MRI pulse sequence programming have been developed, including Pulseq(29), PyPulseq(30), TOPPE(31), and gammaSTAR(32). Among these, Pulseq has been widely adopted because it supports sequence development in MATLAB and Python and provides an open sequence file format that can be interpreted on scanners from different vendors. Pulseq-based implementations have been demonstrated for a range of quantitative and advanced MRI applications, including chemical exchange saturation transfer (CEST)(33), T1 and T2 mapping(34), diffusion MRI(35), and X-nuclear imaging(36). However, to our knowledge, PCASL has not yet been implemented within the Pulseq framework, and whether a shared vendor-neutral PCASL definition can improve the inter-scanner reproducibility of quantitative perfusion measurements has not been systematically evaluated.

We hypothesized that a shared PCASL sequence definition would reduce inter-scanner variability by minimizing differences arising from sequence implementation, such that inter-scanner variability could approach the level of within-scanner test–retest variability. In this study, we developed a vendor-agnostic PCASL sequence using the Pulseq framework and deployed it across 3T MRI systems from three different vendors (Siemens, GE, and United Imaging). By enabling an identical sequence definition to be executed across platforms, this work addresses a key barrier to quantitative ASL imaging in multi-center studies, namely the lack of acquisition-level standardization. We evaluated within-scanner repeatability and inter-scanner reproducibility across the three vendor platforms(18), with additional comparison to a vendor-based PCASL sequence. The framework was further extended to support multi-post-labeling delay (multi-PLD) PCASL for simultaneous estimation of CBF and ATT, as well as vessel-encoded PCASL (VE-PCASL)(37) for territorial perfusion imaging using a planning-free strategy that avoids the need for an angiographic scout and simplifies the acquisition workflow(38). Preliminary results from this work were presented at the 2026 ISMRM Annual Meeting(39). By providing a unified and extensible implementation of PCASL across scanner platforms, this work establishes an acquisition-level approach to sequence standardization and provides a basis for improving the reproducibility and comparability of quantitative ASL measurements in multi-center studies.

# 2 Methods

## 2.1 Development of PCASL in Pulseq framework

The PCASL sequence was developed in MATLAB using Pulseq and deployed across three MRI platforms (Siemens, GE, and United Imaging), with conversion to TOPPE for execution on the GE system (Figure 1a)(31,36). The sequence was organized into four independently modifiable modules: pre-saturation, labeling/control, background suppression, and 2D echo-planar imaging (EPI) readout. The modular design and corresponding complete sequence diagram are illustrated in Figure 1b.

**Pre-saturation:** A $B_1$-insensitive WET-based pre-saturation module consisting of four Gaussian RF pulses (flip angles: 88.9°, 98.7°, 82.5°, 159°; total duration: 7.68 ms) was used to suppress longitudinal magnetization over a broad $T_1$ range (40,41).

**Labeling/Control module:** A balanced PCASL labeling scheme was implemented using a train of Gaussian RF pulses (flip angle: 20°, duration: 0.5 ms, interval: 1 ms), applied concurrently with a z-gradient ( $G_{max}$: 6 mT/m, $G_{av}$: 0.8 mT/m). The phase increment between successive RF pulses in the labeling train was defined as

$$\Delta \phi = \gamma \cdot G_{av} \cdot \Delta z \cdot \Delta t \, ,$$

where $\Delta z$ is the offset of the labeling plane, and $\Delta t$ is the pulse interval. The labeling plane was positioned 90 mm inferior to the center of the imaging slab. The control condition was generated by applying an additional 180° phase shift to every second RF pulse.

For planning-free VE-PCASL, the labeling module was extended by applying three additional sets of bipolar gradients along the x and y directions during the labeling train. An additional phase shift of 90° or 270° was applied to every second RF pulse according to the encoding direction(38).

**Background suppression:** Two frequency offset corrected inversion (FOCI) pulses(42,43)($\beta$ = 764 rad/s, μ = 7.06, max $B1^{+}$ = 20 μT, duration = 10.24 ms) were placed during the post-labeling delay (PLD) period to suppress the signal of the static tissue. The inversion timings were calculated based on the optimized $T_1$ value ($T_{1,opt}$) and the time from the end of the pre-saturation to the image readout (TI)(44):

$$BS_1 = TI + 2T_{1,opt} \cdot \ln\left(0.25 + 0.75\, e^{-\frac{TI}{2T_{1,opt}}}\right)$$

$$BS_2 = TI + 2T_{1,opt} \cdot \ln\left(0.75 + 0.25\, e^{-\frac{TI}{2T_{1,opt}}}\right)$$

By default, we set $T_{1,opt}$ = 500 ms, but the value may be reduced when necessary to ensure the first background suppression pulse was positioned after the labeling or control module.

**Readout:** The readout module was implemented using a 2D EPI readout with ramp sampling. Phase correction was enabled by acquiring three navigator echoes with the blip gradient (Gy) turned off at the beginning of each slice.

## 2.2 In vivo experiments

Seven healthy volunteers (males, age 20-25 years) were recruited and gave informed consent. The study protocols were approved by the local institutional review board. Data were acquired on three 3T MRI systems: Prisma (Siemens Healthineers, Erlangen, Germany), Premier (GE HealthCare, Waukesha, WI, USA), and uMR 870 (United

Imaging Healthcare, Shanghai, China). The receive coils used on the three systems were 20-channel, 16-channel, and 34-channel head coils, respectively.

All PCASL acquisitions employed a 2D EPI readout with identical imaging parameters across scanners unless otherwise specified: FOV = 220 × 220 × 120 mm$^3$, spatial resolution = 3.4 × 3.4 × 4.5 mm$^3$, slice spacing = 0.5 mm, interslice time = 45 ms, TE = 14 ms, bandwidth = 2000 Hz/pixel, flip angle = 90°, 6/8 partial Fourier, and no parallel imaging acceleration. An $M_0$ calibration image was acquired using the same readout without pre-saturation, labeling, or background suppression. Vendor-specific adjustments were implemented at the system configuration level to accommodate hardware differences.

Single-PLD PCASL acquisitions were performed with a labeling duration (LD) of 1.8 s, PLD of 1.8 s, repetition time (TR) of 4.8 s, and 20 label/control pairs (scan time ≈ 3.3 min). Test-retest experiments were conducted in all subjects, with two identical scanning sessions performed on the same day. In each session, data were acquired on all three scanners. On the Siemens MRI system, both the vendor-based PCASL sequence, which was adapted in-house from the vendor's product pulsed ASL sequence on the Siemens IDEA platform, and the Pulseq-based PCASL sequence were acquired sequentially, whereas only Pulseq-based acquisitions were performed on the GE and UIH systems. This design yielded four distinct sequence implementations per session (Vendor on Siemens, and Pulseq on Siemens, GE, and UIH), enabling assessment of both intra-scanner repeatability (test–retest) and inter-scanner reproducibility against the vendor baseline.

To further demonstrate the flexibility of the proposed framework, two additional PCASL sequence variants were evaluated. A multi-PLD PCASL protocol was acquired in one healthy volunteer (age 24 years) to assess arterial transit time (ATT). Six PLDs (0.25, 0.50, 0.75, 1.00, 1.25, and 1.50 s) were selected to cover the expected gray matter ATT range, with images acquired in separate blocks at each PLD. Seven label/control pairs were collected per PLD, with TR set to the minimum achievable value, resulting in a total scan time of approximately 6 minutes.

A planning-free VE-PCASL sequence was acquired in one healthy volunteer across all three scanners to evaluate cross-vendor feasibility. All acquisition parameters followed those of the single-PLD experiment, except that the standard labeling module was replaced with the VE-PCASL implementation. Four labeling configurations were acquired in separate dynamic blocks: a nonselective condition ("label-all") and three vessel-encoded conditions (RL, AP1, and AP2). Images were acquired in a fixed cycle of control–label–RL–AP1–AP2, repeated 20 times, giving a total scan time of approximately 8 minutes.

Additionally, a 3D $T_1$-weighted structural image was acquired once per subject on each of the three scanners to generate brain masks (FOV = 220 × 220 × 120 mm$^3$, 1 × 1 × 1 mm$^3$ isotropic resolution).

## 2.3 Data processing

All PCASL data were reconstructed using a custom MATLAB pipeline, including ramp-sampling interpolation, linear Nyquist ghost correction based on three navigator echoes, and Fourier reconstruction. The resulting images were processed using oxasl(45) for motion correction, EPI distortion correction, and pairwise label–control subtraction to generate perfusion-weighted images (PWI). For multi-PLD PCASL, in-plane Gaussian smoothing (SD = 2.5 mm) was additionally applied. For VE-PCASL, each encoded image (label-all, RL, AP1, and AP2) was subtracted from its corresponding control image to obtain encoded PWI. Vascular territory maps were generated using k-means clustering for initial territory identification, followed by voxel-wise linear fitting to estimate individual feeding artery contributions (38).

For the single-PLD PCASL, the 3D $T_1$-weighted structural images were processed with FSL(46) (FMRIB Software Library, Oxford) for brain extraction(47) and partial volume estimation for cerebrospinal fluid (CSF), gray matter (GM), and white matter (WM)(48). These partial volume maps were down-sampled to the PCASL resolution, and thresholds of 5% for brain tissue, 50% for GM, and 80% for WM were applied to generate whole-brain, GM, and WM masks. For the multi-PLD PCASL and VE-PCASL, whole-brain masks were generated from the M0 images, manually edited using FSLeyes. The $M_0$ calibration images and PWI were masked for subsequent analysis. Quantitative CBF maps were obtained for both single-PLD and multi-PLD PCASL using oxasl (49,50) based on the standard PCASL kinetic model (5), with ATT maps additionally estimated for multi-PLD PCASL. For voxel-wise comparisons, the single-PLD CBF maps and corresponding tissue masks were registered to the $T_1$-weighted structural images and normalized to Montreal Neurological Institute (MNI) space using SPM12 (51).

Region-based and voxel-based comparisons were performed to evaluate the reproducibility and agreement among four implementations. For the subject-wise comparison, the $CV_{within}$, root mean square error (RMSE), bias, and linear correlation of the regional CBF were calculated across seven subjects. For test-retest comparisons, $CV_{within}$ and RMSE were calculated directly between the two repeated scans. For comparisons between implementations, metrics ($CV_{within}$, RMSE, and bias) were computed using a strict pairwise matching approach(52) to prevent artificial error reduction from noise suppression caused by averaging repeated measurements. Specifically, the differences were calculated independently between the first scans and the second scans, and the resulting error metrics were subsequently averaged.

For each subject i, implementation A, and scan r (r = 1, 2), $C_{i,A,r}$ denotes the mean CBF within the GM or WM ROI. For any two matched ROI mean CBF measurements X and Y, CVwithin and RMSE were defined as:

$$\mathrm{CV}_{within}(X,Y) = \frac{\frac{|X-Y|}{\sqrt{2}}}{\frac{(X+Y)}{2}} \times 100\%$$

$$\mathrm{RMSE}(X,Y) = \sqrt{(X-Y)^2}$$

For test-retest comparisons of implementation A, X = $C_{i,A,1}$ and Y = $C_{i,A,2}$. For comparisons between implementation A and B, X = $C_{i,A,r}$ and Y = $C_{i,B,r}$ for each matched scan (r = 1, 2). The two pairwise CVwithin and RMSE values were then averaged for each subject.

For the voxel-wise comparison, CBF and $CV_{within}$ values were computed for each voxel to assess reproducibility differences spatially. In order to visualize how much larger or smaller the between-implementation standard deviation of paired CBF differences ($SD_{\Delta CBF}$) was compared to the test-retest $SD_{\Delta CBF}$, a voxel-wise variation ratio map was created according to the following equation(16):

$$\text{Variation Ratio} = 100\% \frac{SD_{\Delta CBF,betwenn-inplementation}}{SD_{\Delta CBF,test-retest}}$$

For each Pulseq implementation, the ratio compares the standard deviation of paired CBF differences relative to Siemens-Vendor with the standard deviation of test–retest differences. The Siemens-Pulseq comparison is within the same scanner; the GE-Pulseq and UIH-Pulseq comparisons are inter-scanner. If the variance between implementations is equal to the intra-scanner test-retest variance, a mean variation ratio of 100% is expected. Finally, individual CBF histograms were averaged to generate a group-level histogram. Corresponding GM $CV_{within}$ and variation ratio histograms were also generated to further quantify the spatial error distributions.

# 3 Results

## 3.1 Region-based comparison

Figure 2 presents six representative slices of PWI acquired from the same subject using three Pulseq-based PCASL implementations on Siemens, GE, and UIH scanners, together with a vendor-based PCASL sequence on Siemens. To facilitate a direct comparison of the reconstructed perfusion signals, all images were generated using raw data reconstruction and pairwise label–control subtraction, without additional post-processing or distortion correction. Despite minor geometric distortions associated with EPI susceptibility effects, all four implementations produced visually comparable perfusion contrast, spatial distributions, and overall image quality, with no substantial differences apparent in the representative images.

Figure 3 compares the quantitative CBF maps and regional measurements obtained from the vendor-based implementation and the three Pulseq-based acquisitions across seven subjects. The resulting CBF maps showed consistent spatial perfusion patterns, with preserved gray–white matter differentiation across platforms. The mean GM CBF values (mL/100g/min) were 52.51 ± 6.43 for Siemens-Vendor, 53.64 ± 6.64 for Siemens-Pulseq, 52.25 ± 6.84 for GE-Pulseq, and 52.41 ± 5.37 for UIH-Pulseq, demonstrating close agreement in absolute perfusion estimates. Relative to Siemens-Vendor, Siemens-Pulseq showed a small positive GM CBF bias of 1.13 ± 1.39 mL/100g/min, whereas the mean signed WM CBF bias was close to zero (−0.18 ± 1.19 mL/100g/min). These results indicate that the Pulseq-based implementations yielded comparable regional CBF estimates across the three vendor platforms (52.25–53.64 mL/100g/min), with differences of no more than 1.4 mL/100g/min between platforms.

Figure 4 presents scatter plots of region-wise GM and WM CBF measurements for test–retest (Figure 4a) and comparisons between implementations (Figure 4b). All four implementations demonstrated strong test–retest correlations, with GM correlations ranging from R = 0.850 to 0.988 and WM correlations from R = 0.813 to 0.961. Siemens-Pulseq exhibited the highest test–retest correlations (GM: R = 0.988; WM: R = 0.961), compared with Siemens-Vendor (GM: R = 0.891; WM: R = 0.829), GE-Pulseq (GM: R = 0.945; WM: R = 0.820), and UIH-Pulseq (GM: R = 0.850; WM: R = 0.813). Comparisons of Siemens-Vendor with each of the three Pulseq implementations showed strong across-subject correlations in both GM (R = 0.967–0.981) and WM (R = 0.790–0.942). Specifically, the GM/WM correlations were 0.977/0.942 for Siemens-Pulseq, 0.981/0.936 for GE-Pulseq, and 0.967/0.790 for UIH-Pulseq. Pairwise comparisons among the three Pulseq implementations further demonstrated consistent cross-platform associations, with GM correlations ranging from 0.934 to 0.978 and WM correlations from 0.802 to 0.958. Overall, GM CBF measurements exhibited consistently stronger correlations than WM measurements, and all inter-scanner correlations were statistically significant (GM: $P < 0.001$; WM: $P < 0.05$). These results demonstrate that the regional variation in GM and WM CBF was consistently reproduced across implementations and scanner platforms.

Beyond reproducing the relative perfusion pattern, the absolute CBF values also showed close numerical agreement across implementations (Figure 5). GM CBF $CV_{within}$ remained below 4% for all four implementations, with Siemens-Pulseq yielding the lowest value (2.65 ± 1.39%), compared with Siemens-Vendor (3.64 ± 3.42%); GE-Pulseq and UIH-Pulseq fell within a similar range (3.50 ± 1.48% and 3.28 ± 2.40%, respectively). WM $CV_{within}$ values were comparable between Siemens-Pulseq and Siemens-Vendor (4.70 ± 2.19% vs. 4.77 ± 3.47%), indicating similar measurement variability between the Pulseq and vendor-based implementations. Similarly, the intra-scanner test-retest GM RMSE remained below 3 mL/100g/min for all platforms (Siemens-Vendor: 2.53 ± 2.13; Siemens-Pulseq: 1.97 ± 1.05; GE-Pulseq: 2.68 ± 1.34; UIH-Pulseq: 2.51 ± 1.77 mL/100g/min), demonstrating consistently low absolute test–retest errors across scanner platforms and implementations. Furthermore, GM RMSE for Siemens-Vendor compared with each Pulseq implementation (1.96 ± 0.81 for Siemens-Pulseq, 1.64 ± 0.80 for GE-Pulseq, and 1.93 ± 0.80 for UIH-Pulseq mL/100g/min) fell within the same range as these test-retest values, with all corresponding $CV_{within}$ remaining below 3%. Importantly, the inter-scanner differences were comparable to the intra-scanner test–retest variability, supporting the quantitative consistency of Pulseq-based PCASL across vendors.

## 3.2 Voxel-based comparison

Figure 6 presents the group-averaged voxel-wise CBF maps and corresponding CBF histograms across all four implementations, derived by spatially normalizing individual CBF maps to a common template and averaging across subjects. Spatially, all implementations preserved clear gray–white matter differentiation across cortical and subcortical regions, with broadly consistent perfusion patterns across views. Siemens-Pulseq and UIH-Pulseq showed high spatial correspondence with Siemens-Vendor, though UIH-Pulseq exhibited slightly reduced perfusion in the frontal lobe, and Siemens-Pulseq showed a somewhat stronger GM CBF signal in the posterior flow territory. GE-Pulseq, by contrast, displayed noticeably reduced signal in the posterior flow territory alongside relatively elevated perfusion in the parietal lobe, suggesting a platform-specific difference in the spatial distribution of CBF. These spatial observations were supported by the CBF histograms, in which Siemens-Vendor, Siemens-Pulseq, and UIH-Pulseq all exhibited broadly similar bimodal distributions reflecting the GM–WM contrast — with only a slight leftward shift of the first peak in Siemens-Pulseq and a mild rightward shift of the second peak in UIH-Pulseq — whereas the GE-Pulseq histogram showed the two peaks converging into a single broader distribution, consistent with the reduced posterior perfusion contrast observed in the CBF maps.

Figure 7 illustrates the voxel-wise test–retest variability across all acquisitions. The $CV_{within}$ maps and histograms were broadly similar, with overall variability remaining lower than that reported in a previous study(16). Notably, Siemens-Pulseq exhibited a variability profile nearly identical to that of Siemens-Vendor. In contrast, GE-Pulseq showed higher variability in the posterior flow territory, whereas UIH-Pulseq demonstrated elevated variability in the frontal and parietal lobes. These regional differences were reflected in the corresponding histograms, with both GE-Pulseq and UIH-Pulseq exhibiting a slight rightward shift and broader distribution of $CV_{within}$ values relative to Siemens-Vendor and Siemens-Pulseq.

The ratio of between-implementation to test–retest variability (Figure 8) was heterogeneously distributed, ranging from 50% to 150% in the majority of voxels. Histogram analysis revealed unimodal distributions with peaks centered around 100%, indicating that variability between implementations was broadly comparable to test-retest variability across most voxels. Siemens-Pulseq peaked slightly below 100%, whereas UIH-Pulseq peaked slightly above 100%. Compared with Siemens-Pulseq, GE-Pulseq and UIH-Pulseq exhibited broader right-tailed distributions, reflecting a larger proportion of voxels in which the between-implementation $SD_{\Delta CBF}$ exceeded the test-retest $SD_{\Delta CBF}$. Notably, Siemens-Pulseq showed fewer such voxels, likely because Siemens-Pulseq and Siemens-Vendor were acquired consecutively within the same scanning session, which may have limited physiological fluctuations between acquisitions. A minority of voxels exhibited extreme variation ratios exceeding 300%, with their spatial distribution differing across comparisons: in the Siemens-Vendor versus GE-Pulseq comparison, these outliers were predominantly located in the posterior flow territory; in the Siemens-Vendor versus UIH-Pulseq comparison, they were scattered throughout the brain; whereas in the Siemens-Vendor versus Siemens-Pulseq comparison, such extreme outlier voxels were virtually absent.

## 3.3 Extended PCASL implementations

Beyond the single-PLD evaluations, two additional acquisition protocols, multi-PLD PCASL and planning-free VE-PCASL, were implemented and tested on all three vendor platforms.

Figure 9 presents results from the multi-PLD protocol, in which data were acquired at six post-labeling delays to enable simultaneous estimation of CBF and ATT. Quantitative CBF maps (Figure 9a) and ATT maps (Figure 9b) were successfully generated on all three platforms. The CBF maps showed spatially coherent perfusion with preserved GM–WM contrast, and the ATT maps showed regional variations qualitatively consistent with vascular anatomy.

Figure 10 shows results from the planning-free VE-PCASL implementation, which modifies the labeling module to encode vascular territory information without requiring an angiographic scout or manual vessel planning. Images acquired under four encoding configurations (Label-all, RL, AP1, and AP2) showed distinct spatial modulations in the perfusion signal (Figure 10a). Subsequent decoding resolved the territories supplied by the right internal carotid artery (RICA), left internal carotid artery (LICA), and basilar artery (BA), with broadly consistent delineation across the three scanners (Figure 10b). Fine anatomical structures, including the choroid plexus, were also identifiable across platforms. These results demonstrate the feasibility of implementing both extended PCASL protocols within the common Pulseq framework on Siemens, GE, and UIH systems.

# 4 Discussion

In this study, we developed a modular PCASL sequence using Pulseq and evaluated its performance across three 3T MRI systems from Siemens, GE, and UIH. The same sequence was used on all three platforms, yielding closely agreeing mean GM CBF measurements and test–retest variability comparable to that of the Siemens-Vendor reference. WM measurements also showed good repeatability, although correlations were generally lower than those for GM. Despite the overall agreement, voxel-wise analyses revealed localized differences in perfusion and measurement variability. The successful implementation of multi-PLD and planning-free VE-PCASL further demonstrated the applicability of the framework to more advanced ASL acquisitions.

A key feature of the proposed framework is the use of a common PCASL sequence rather than separately developed vendor-specific implementations. Previous studies have demonstrated the feasibility of cross-vendor PCASL using harmonized or near-identical sequences, while also showing that residual differences in sequence parameters can influence CBF quantification(16). In the present implementation, pre-saturation, labeling/control, background suppression, and readout were defined within the same Pulseq sequence, with common RF and gradient waveforms, phase cycling, and event timing. Although execution remains platform-dependent, including TOPPE conversion on GE(29,31), the underlying PCASL sequence does not require independent development in each vendor's programming environment. This approach extends cross-vendor harmonization beyond matching acquisition parameters to standardizing the prescribed sequence itself.

The importance of acquisition-level standardization extends beyond agreement between scanners. Differences in acquisition can affect the interpretation of multi-center studies, as illustrated by the RADAR study, in which variations in readout and PLD introduced artifactual differences in CBF between groups(53). Although known acquisition effects may be accounted for during analysis, consistent sequence implementation reduces potential sources of variability before data collection. This approach is complementary to standardized processing frameworks such as ExploreASL, which harmonize preprocessing, CBF quantification, and quality control (23). Differences in labeling efficiency, background suppression, or readout characteristics may affect the acquired signal in ways that cannot be fully corrected when the underlying effects are unknown. In the present study, all datasets were reconstructed and quantified using a common pipeline, minimizing differences associated with processing choices. Standardization of both acquisition and processing is therefore important for obtaining comparable quantitative ASL measurements across centers.

The quantitative findings showed close agreement in mean GM CBF across the four implementations, ranging from 52.25 to 53.64 mL/100g/min. GM RMSEs relative to the Siemens-Vendor implementation were comparable in magnitude to within-scanner test–retest errors, showing that the observed differences between implementations were of a similar order to those observed in repeated measurements. WM CBF also showed good repeatability, with comparable WM CVwithin for Siemens-Pulseq and Siemens-Vendor (4.70% and 4.77%, respectively). Correlations were generally lower for WM than for GM, consistent with the lower WM perfusion signal and its greater susceptibility to noise and partial-volume effects. However, agreement in mean GM CBF did not extend uniformly throughout the brain. GE-Pulseq showed reduced posterior perfusion and increased voxel-wise variability, whereas UIH-Pulseq exhibited lower frontal perfusion and greater variability in parts of the frontal and parietal lobes. Similar regional discrepancies have been reported in previous cross-vendor PCASL studies despite comparable global GM CBF(18). Although the present study used a common readout design, differences in hardware execution and local labeling conditions may still affect regional measurements. These observations indicate that global CBF agreement should be considered alongside WM and regional perfusion measurements when assessing cross-platform consistency.

Despite the use of an identical sequence definition, differences in hardware response and local field conditions may affect the actual labeling process. Gradient delays and residual eddy currents vary with scanner hardware(54), while RF calibration and RF–gradient timing may influence the realized labeling conditions. In PCASL, gradient imperfections and off-resonance at the labeling plane can perturb the phase relationship between successive RF pulses and flowing spins, reducing labeling efficiency(55,56). The labeling plane was prescribed 90 mm below the imaging slab on all systems, but this fixed offset does not necessarily reproduce the same relationship to the feeding arteries after subject repositioning. Differences in vessel orientation and local B0 conditions may therefore contribute to variations in labeling efficiency(5,57). These factors could partly explain the reduced posterior perfusion observed on GE, although the actual RF and gradient output and labeling-plane B0 were not measured. A common sequence definition thus controls the prescribed acquisition, while hardware characterization and local field calibration remain relevant to achieving consistent labeling performance across platforms.

The modular design also allows the framework to accommodate PCASL methods beyond the single-PLD, 2D EPI acquisition used for quantitative validation. Multi-PLD PCASL was implemented to estimate both CBF and ATT, while planning-free VE-PCASL provided vascular territory information without requiring an angiographic scout or manual vessel-specific encoding prescription(6,38). Both methods were successfully deployed across the three systems by adapting the relevant acquisition modules. Other PCASL implementations could be developed in a similar manner, including alternative readouts such as 3D GRASE or 3D fast spin-echo stack-of-spirals (58,59), as well as different labeling and background suppression strategies(60). Such modifications would require corresponding adjustments to acquisition timing and reconstruction, followed by cross-platform validation. The framework therefore allows different PCASL protocols to be developed within a common sequence architecture while maintaining consistent implementation across vendors.

Beyond the immediate quantitative evaluation, a common sequence implementation may facilitate the development and dissemination of advanced ASL methods. Pulseq provides an open sequence format in which RF and gradient events and their timing can be inspected, shared, and modified independently of vendor-specific programming environments(29). This reduces the need to recreate the same acquisition separately for each scanner platform and allows methodological updates to be distributed through a common implementation. For multi-center and longitudinal studies, sequence files should be accompanied by generation code, interpreter versions, scanner configuration, and reconstruction and quantification settings to support reproducibility and subsequent validation. Such documentation is particularly important when scanner software or hardware is upgraded, since an unchanged sequence file does not necessarily guarantee unchanged acquisition performance.

Several limitations should be acknowledged. A vendor-based PCASL reference was available only on Siemens, so whether Pulseq reduces cross-vendor variability compared with vendor-based implementations remains to be determined. In the current Pulseq implementation, the labeling plane could not be prescribed independently and was instead positioned 90 mm below the imaging slab. This fixed offset may have led to differences in its position relative to the feeding arteries across scans; independent labeling-plane prescription could address this constraint in future implementations.

Labeling efficiency was not measured directly, and agreement across platforms does not necessarily imply accurate CBF quantification. Finally, reconstruction was performed offline, and an online workflow is needed for routine clinical use. An interoperable framework such as Gadgetron could support this transition while allowing a common reconstruction and post-processing pipeline to be deployed across vendors (61).

## 5 Conclusions

This work demonstrates acquisition-level harmonization of PCASL: the same Pulseq sequence yielded mean GM CBF values in close agreement across Siemens, GE, and UIH systems, with between-implementation differences comparable in magnitude to within-scanner test–retest variability. Combined with standardized processing, this approach offers a path toward an end-to-end standardized quantitative ASL workflow, from perfusion signal generation to CBF estimation. Its open and modular design provides a shared foundation for developing, evaluating, and disseminating advanced ASL methods across vendors, supporting reproducible quantitative perfusion imaging in multi-center research.

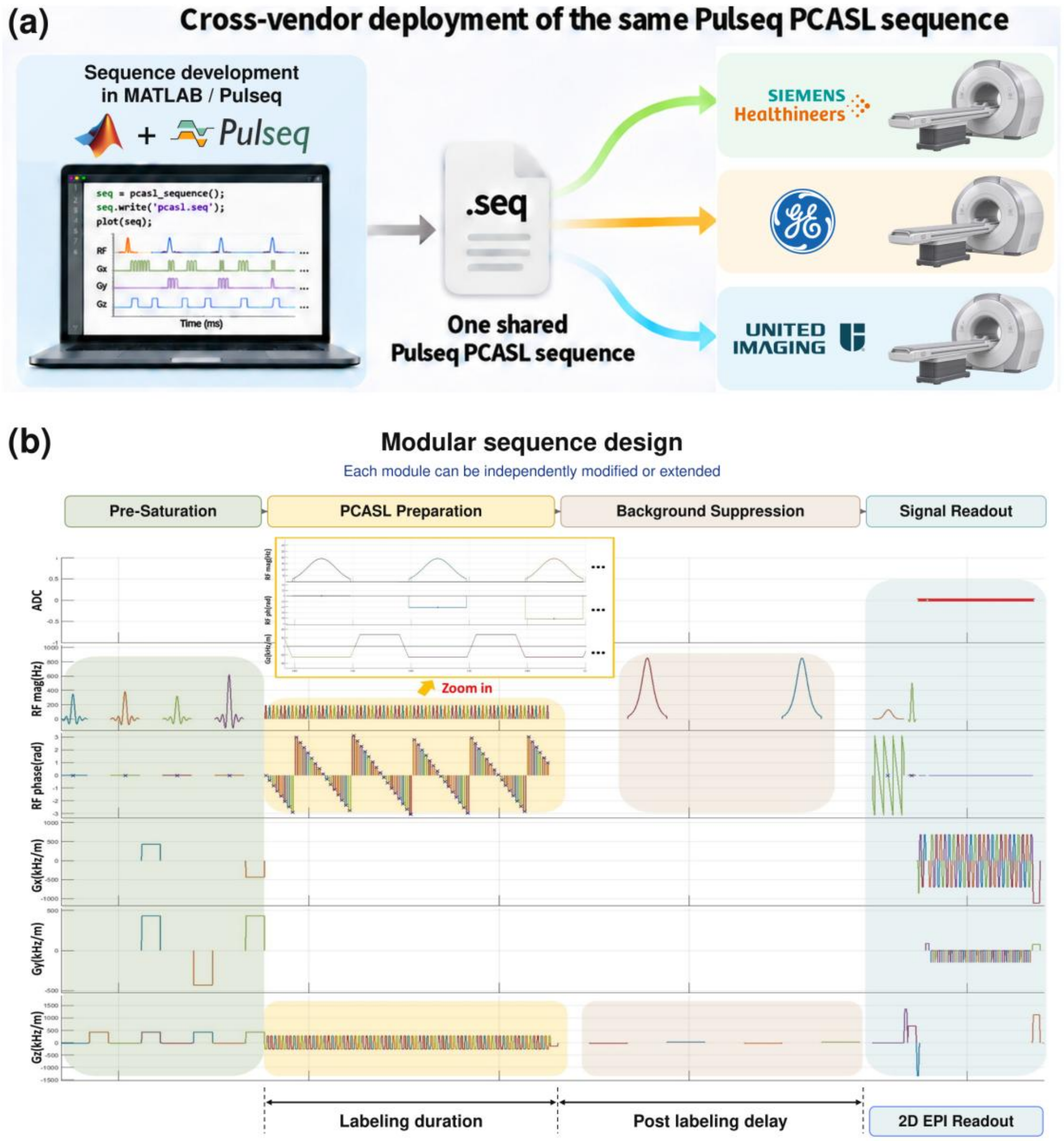

**Figure 1.** Cross-vendor sequence design workflow and modular PCASL architecture. (a) A single PCASL sequence is developed in MATLAB/Pulseq and deployed on Siemens Prisma, GE Premier, and UIH uMR 870 scanners. (b) The sequence comprises four functional modules: pre-saturation, PCASL labeling/control, background suppression, and readout. The modular architecture allows different modules to be substituted independently.

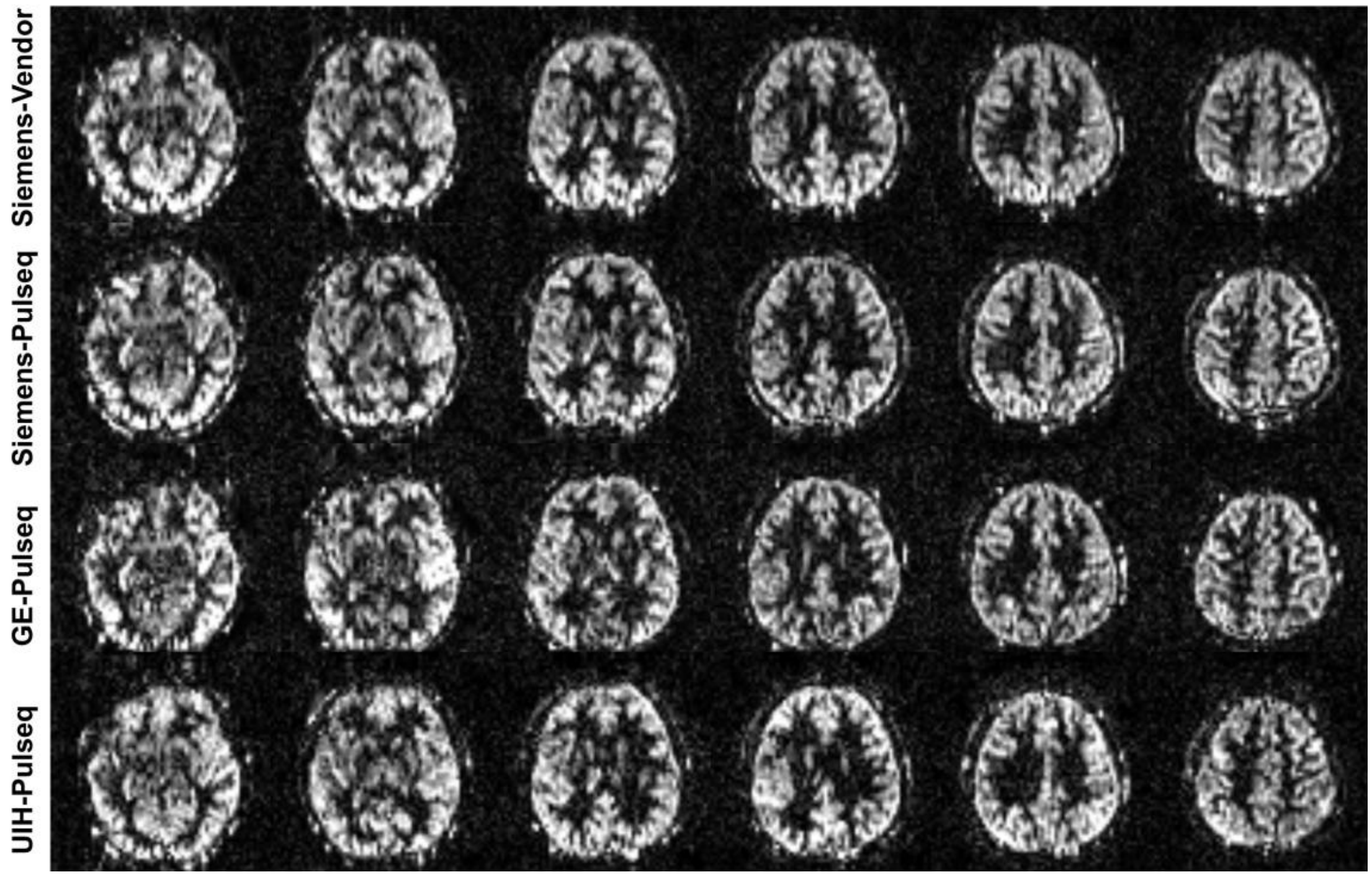

**Figure 2.** Comparison of six perfusion-weighted images (PWI) slices from the same subject acquired using four different implementations on three scanners.

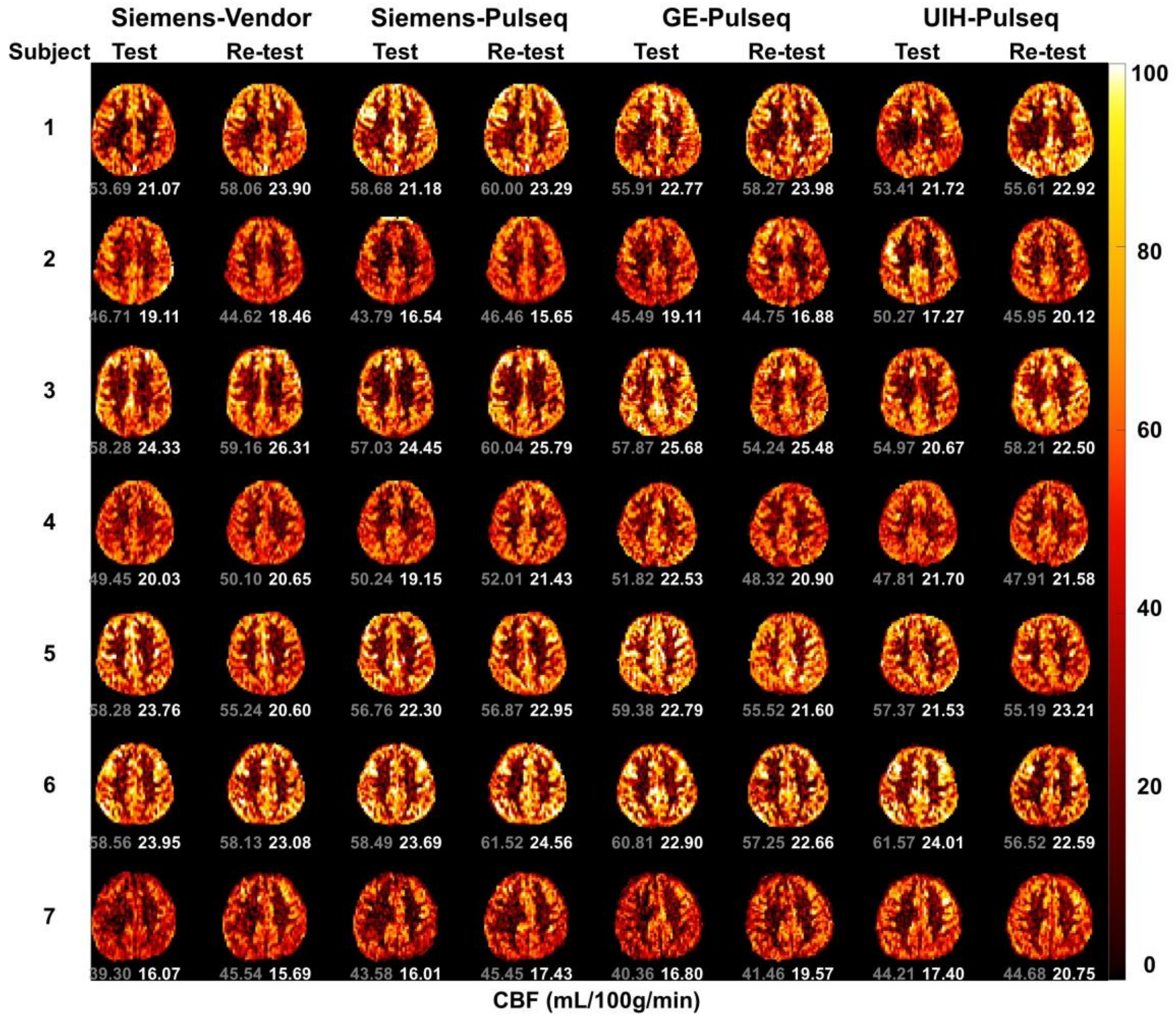

**Figure 3.** A single slice of the cerebral blood flow (CBF) maps for each subject for the comparison between implementations. The mean whole brain GM (gray text) and WM (white text) CBF estimates are inset below each image. GM, gray matter; WM, white matter.

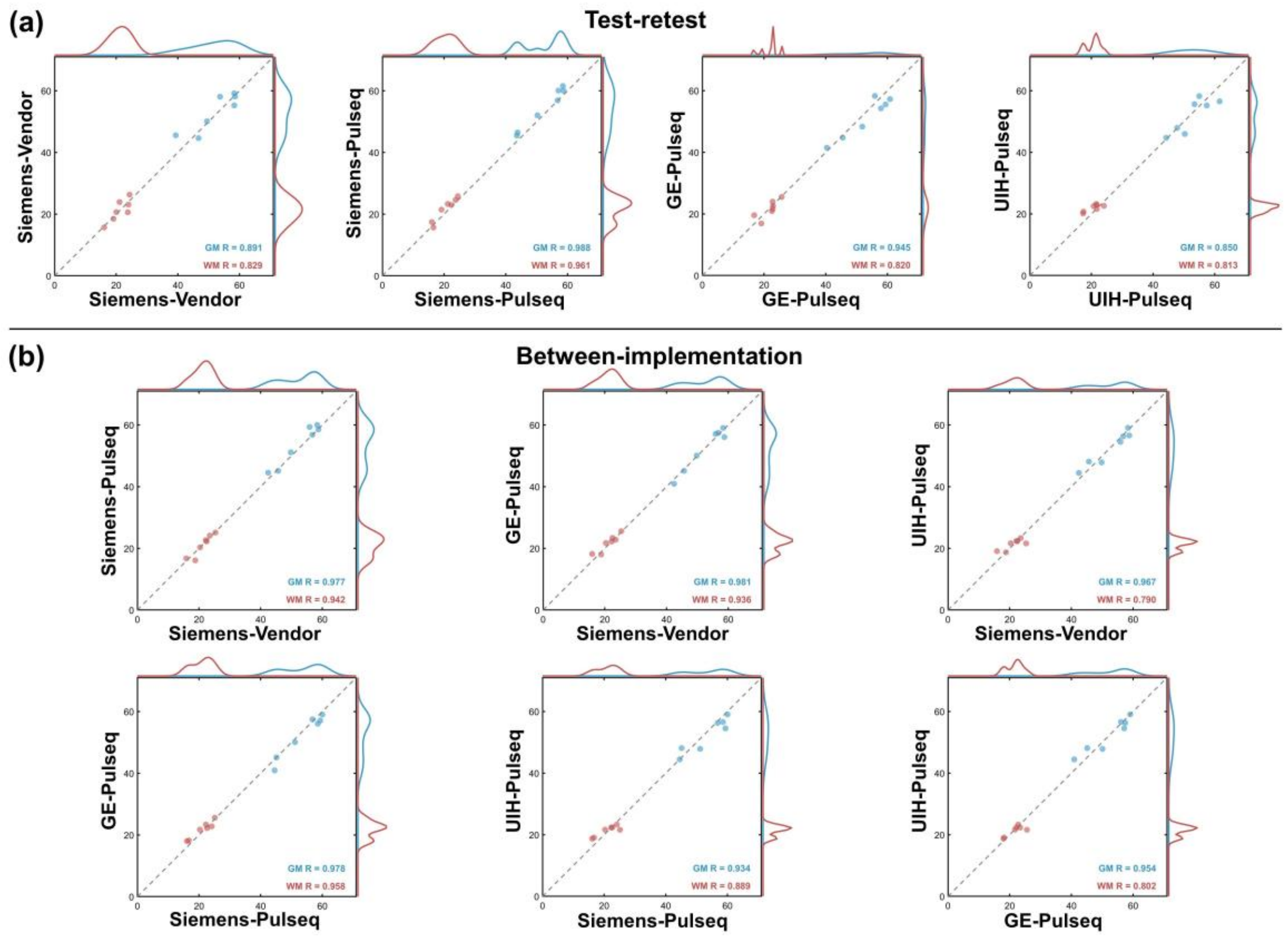

**Figure 4.** Scatter plots of the in vivo region-wise mean cerebral blood flow (CBF) measurements,showing the association of subject-level GM and WM ROI mean CBF across scans and implementations. The graphs show (a) the test-retest comparisons for each implementation and (b) comparisons between Siemens-Vendor and the three Pulseq implementations. Mean gray matter (GM) values are shown in blue and mean white matter (WM) values are shown in red. All correlations between implementations were significant (WM: $P < 0.05$, GM: $P < 0.001$).

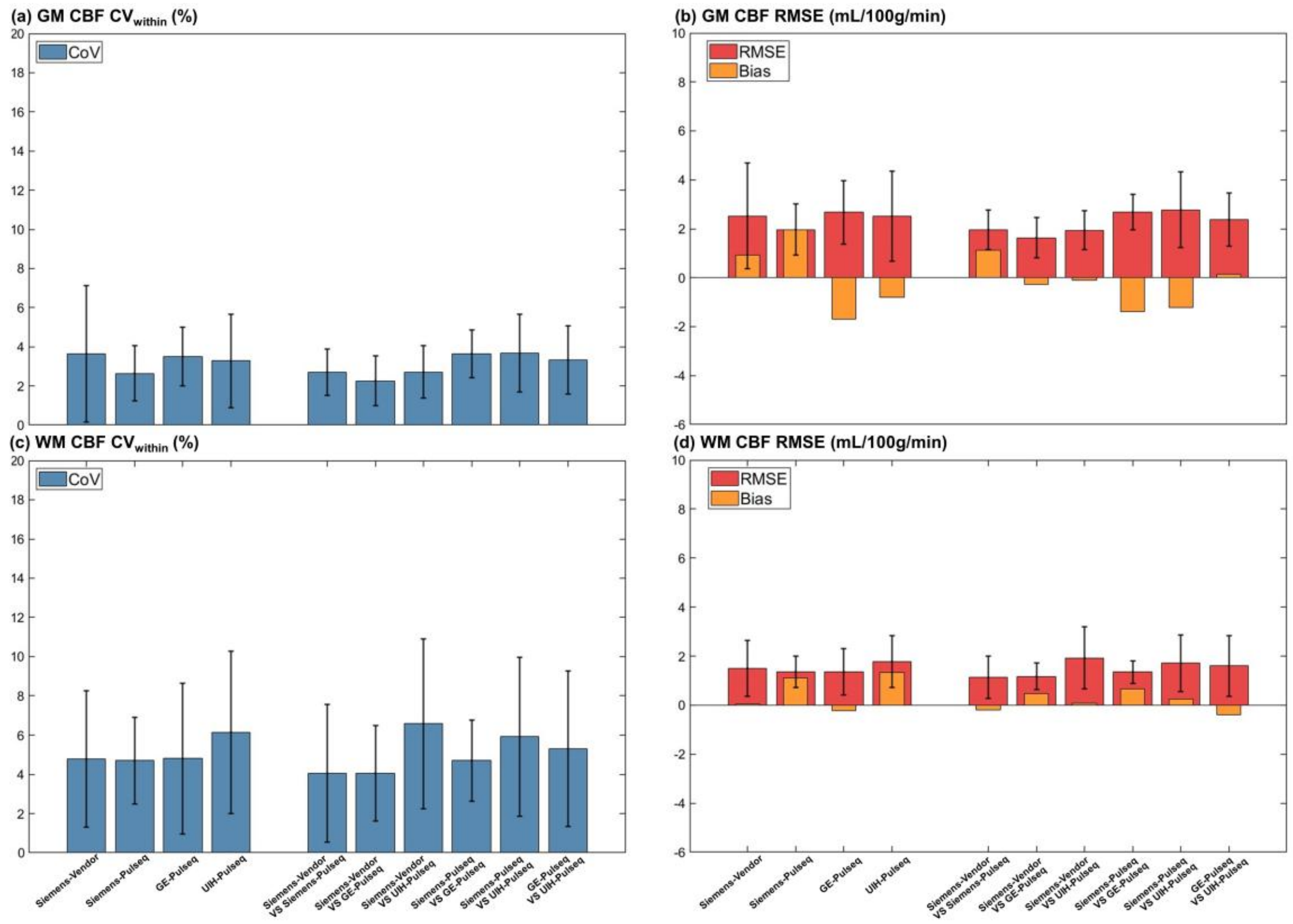

**Figure 5.** Quantitative regional test–retest errors and differences between implementations in cerebral blood flow (CBF) measurements, assessing the magnitude of numerical differences between implementations and platforms. The top-left quadrant shows gray matter (GM) CBF within-subject coefficient of variation ($CV_{within}$); the top-right quadrant shows GM CBF root mean square error (RMSE) and bias; the bottom-left quadrant shows white matter (WM) CBF $CV_{within}$; and the bottom-right quadrant shows WM CBF RMSE and bias. The $CV_{within}$ is a normalized error, whereas the RMSE and bias are in physiological units (mL/100g/min). Bias is signed and indicates the direction of the difference. $CV_{within}$ and RMSE bars show the mean across subjects, with error bars indicating the across-subject standard deviation; bias bars show the mean signed difference without error bars. To prevent artificial error reduction, metrics between implementations were computed using a strict pairwise matching approach. The bias was calculated as mean (B-A), for example, mean (Siemens-Pulseq– Siemens-Vendor) for the Siemens-Vendor versus Siemens-Pulseq comparison.

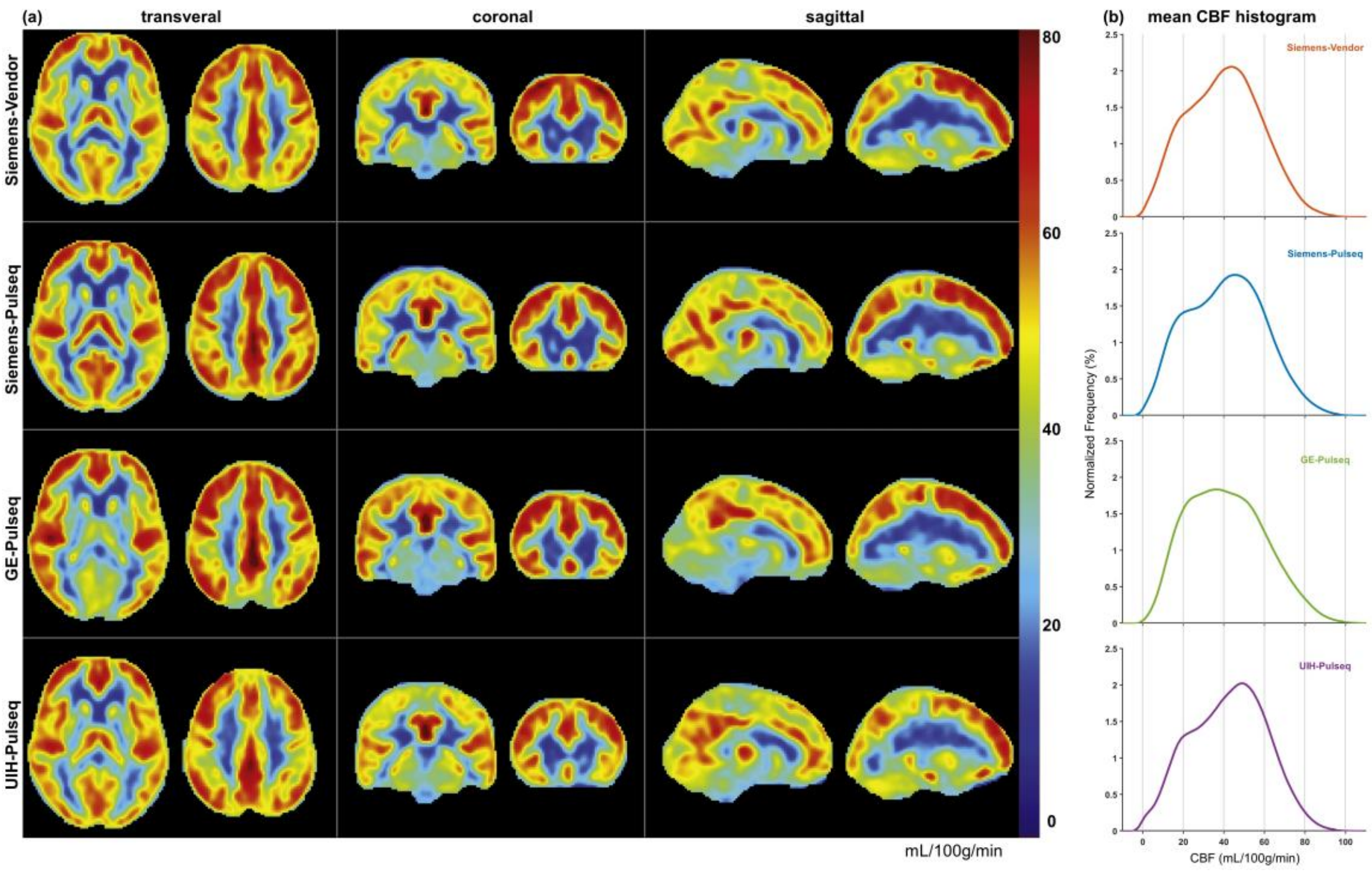

**Figure 6.** Voxel-wise comparison of group-averaged cerebral blood flow (CBF) maps between implementations. (a) CBF maps averaged across subjects following spatial normalization to a common template. (b) Normalized frequency histograms of the group-averaged CBF distributions.

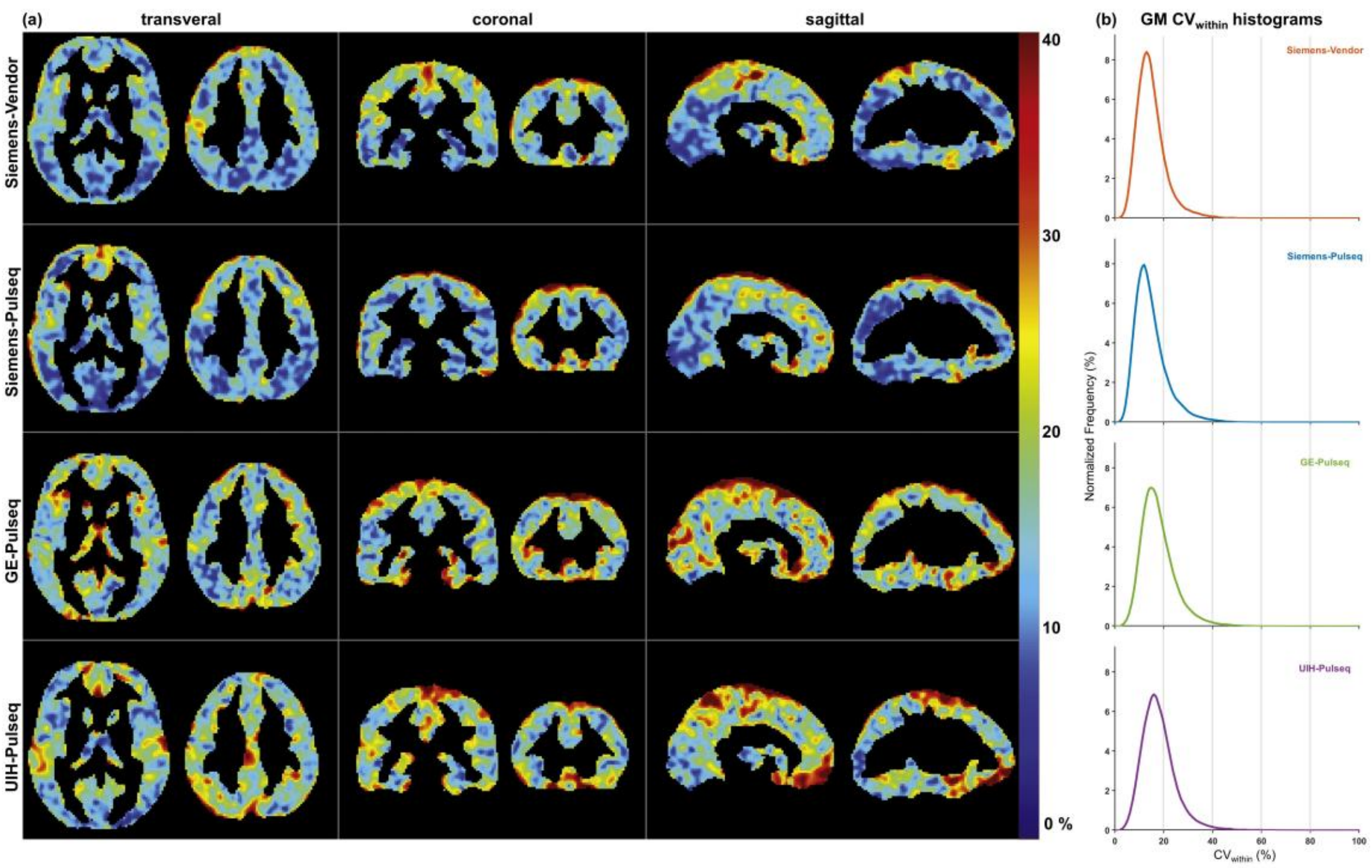

**Figure 7.** Voxel-wise assessment of test-retest reliability using the within-subject coefficient of variation ($CV_{within}$) for different implementations. (a) $CV_{within}$ maps with reference gray matter (GM) mask; (b) Normalized frequency histograms of the GM $CV_{within}$.

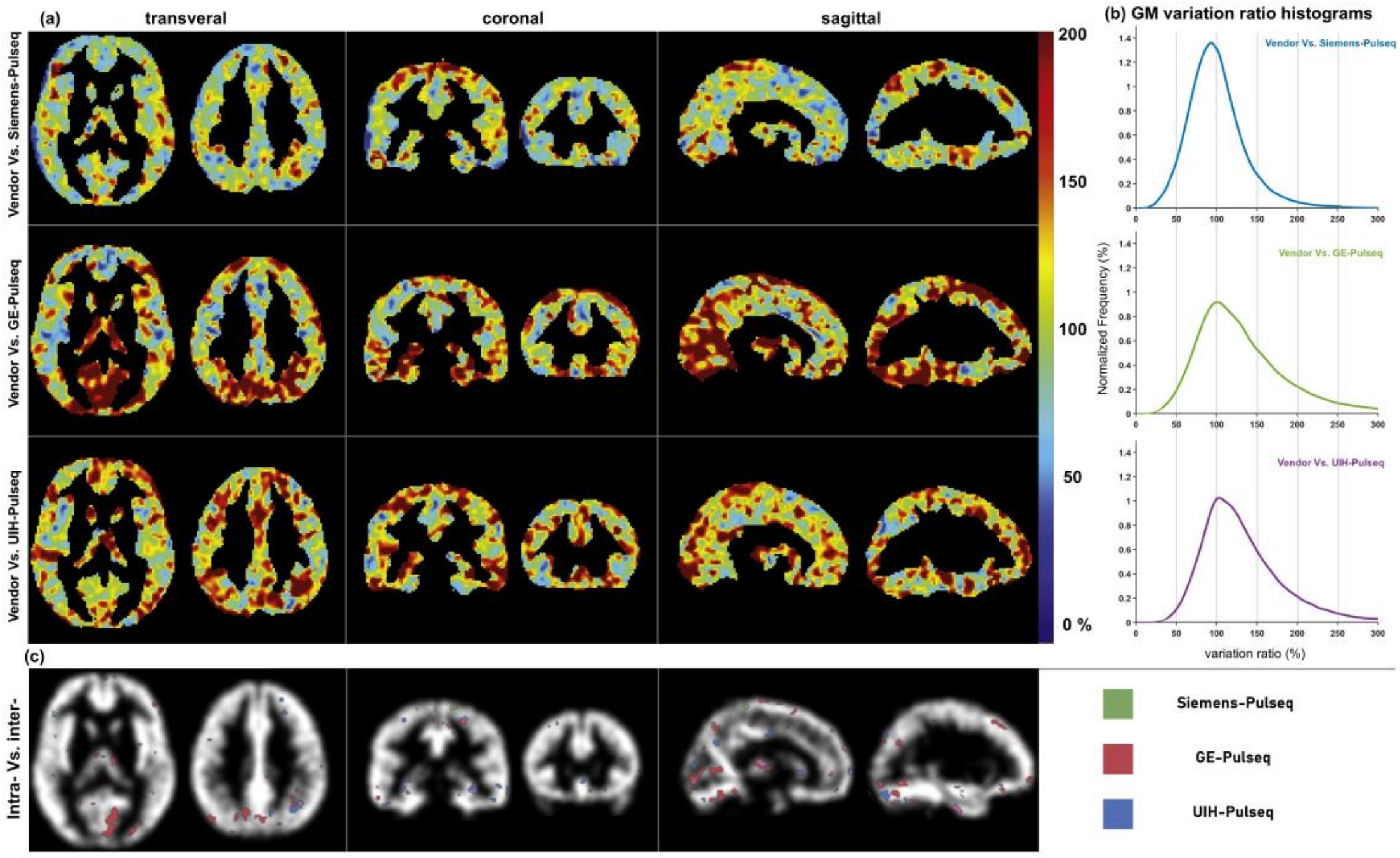

**Figure 8.** Voxel-wise variation ratio demonstrating the relative magnitude of between-implementation versus test–retest variability. (a) Spatial distribution of the variation ratio, defined as the ratio of the between-implementation $SD_{\Delta CBF}$ to the test–retest $SD_{\Delta CBF}$. (b) Normalized frequency histograms of the variation ratio within the gray matter (GM) mask. (c) Binary outlier maps overlaid on mean GM probability map, highlighting voxels with variation ratios exceeding 300%.

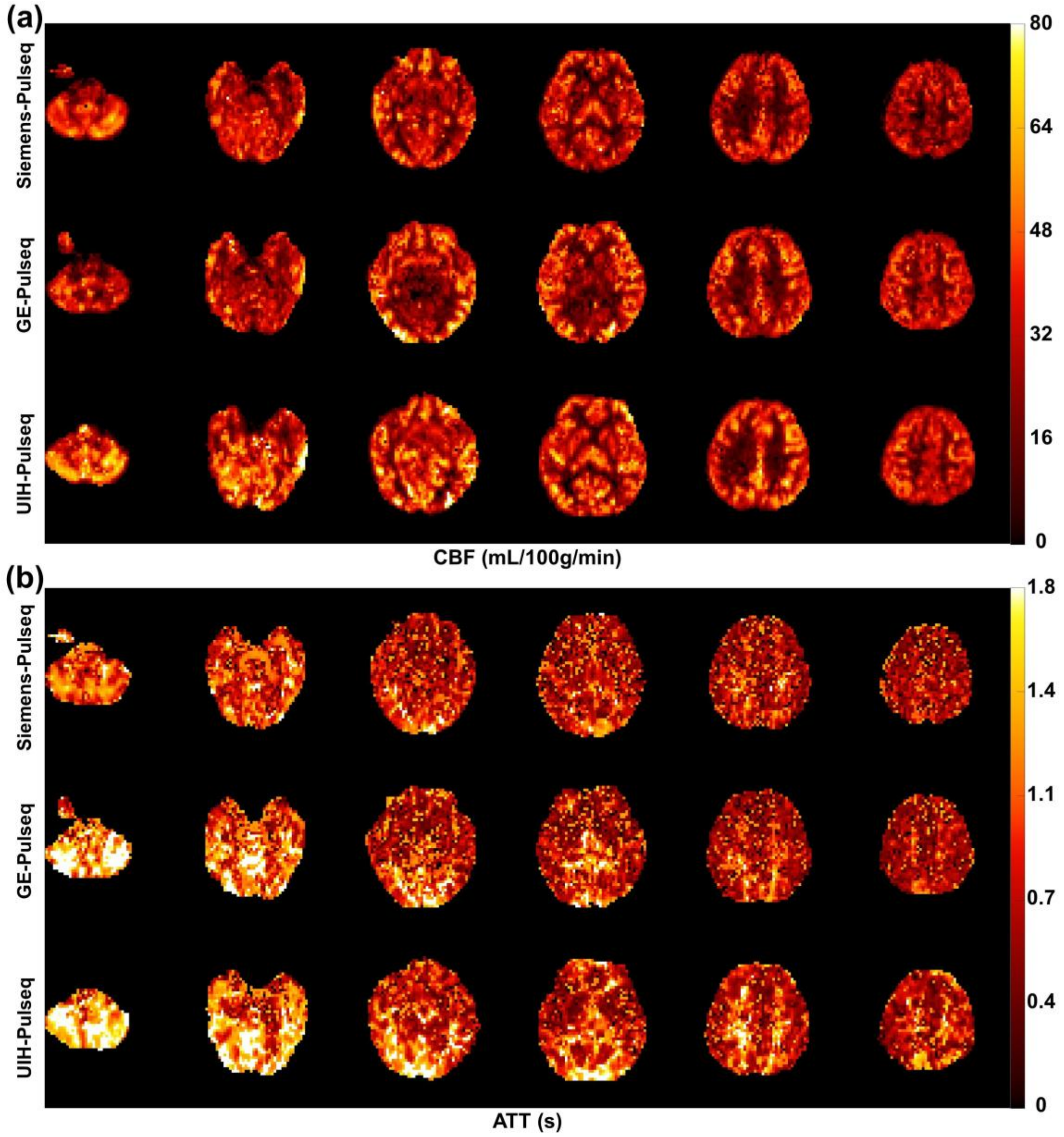

**Figure 9.** Representative quantitative images of six slices from the same subject acquired using the Pulseq-based multi-PLD PCASL sequence across different scanners. (a) Cerebral blood flow (CBF) maps. (b) Arterial transit time (ATT) maps.

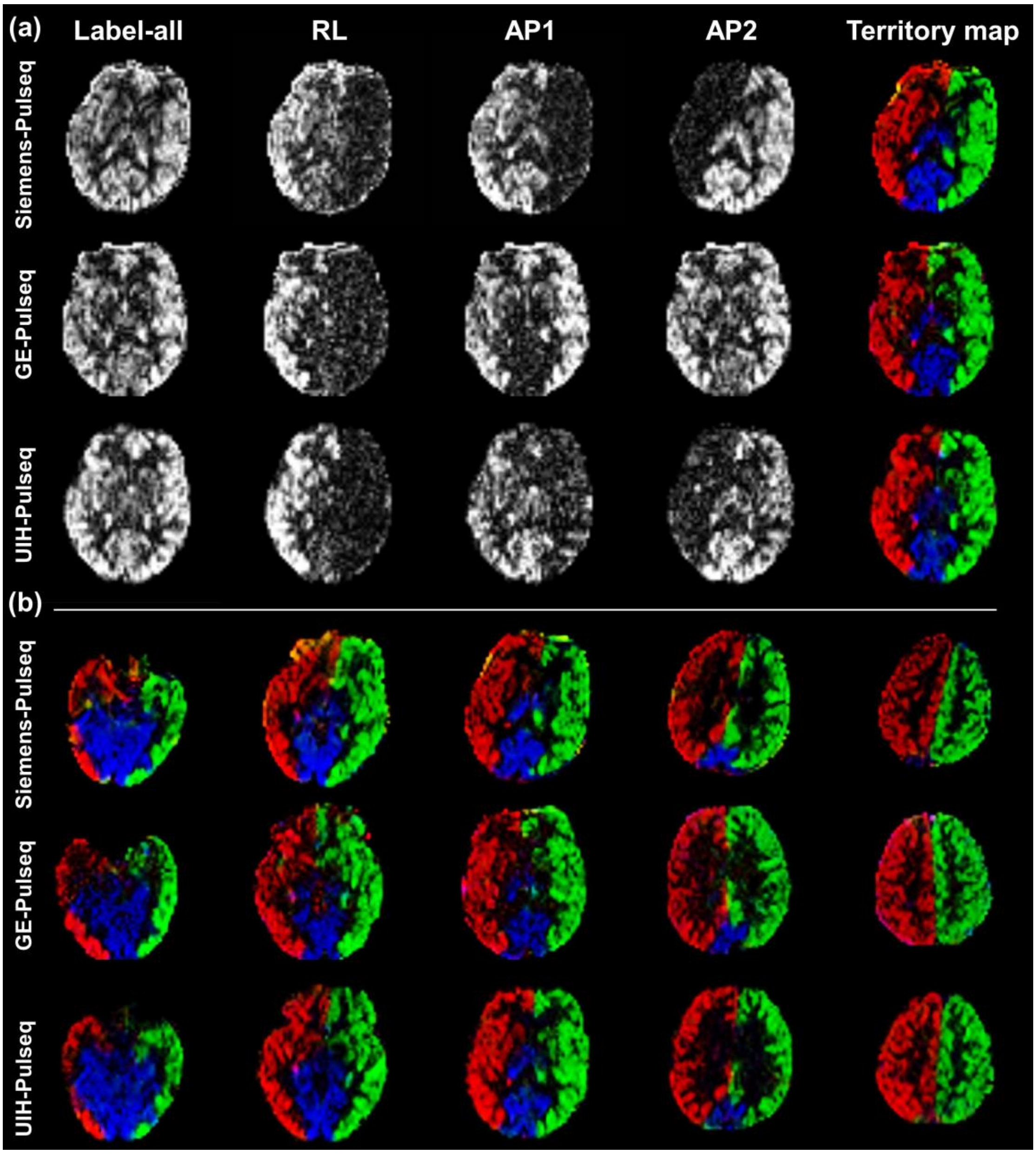

**Figure 10.** Vascular territory maps of planning-free vessel-encoded PCASL in the Pulseq framework. (a) Perfusion-weighted images (PWI) from a single slice acquired with different vessel-encoding patterns (Label-all, RL, AP1 and AP2) and corresponding decoded territory maps. (b) Five representative slices of vascular territory maps segmented into right internal carotid artery (RICA), left internal carotid artery (LICA), and basilar artery (BA).